\documentclass[doublecol]{epl2} 
\usepackage{color}
\usepackage{soul}
\usepackage{amsmath}
\DeclareMathOperator*{\Max}{Max}
\title{Dynamical self-stabilization of the Mott insulator: Time evolution
of the density and entanglement  entropy of out-of-equilibrium cold fermion gases}
\shorttitle{Dynamical self-stabilization of the Mott insulator} 

 \author{Daniel Karlsson \inst{1} \and Claudio Verdozzi \inst{1} \and Mariana M. Odashima\inst{2} \and Klaus Capelle\inst{3}}
\shortauthor{D. Karlsson \etal}

\institute{                    
 \inst{1} Mathematical Physics and European Theoretical Spectroscopy Facility,
Lund University, 22100 Lund, Sweden\\
 \inst{2} Instituto de F\'{\i}sica de S\~ao Carlos,
Universidade de S\~ao Paulo, S\~ao Carlos, 13560-970 S\~ao Paulo,
Brazil\\
\inst{3} Centro de Ci\^encias Naturais e Humanas,
Universidade Federal do ABC, Santo Andr\'e, 09210-170 S\~ao Paulo,
Brazil
}
\pacs{37.10.Jk}{Atoms in optical lattices}
\pacs{71.15.Mb}{Density functional theory, local density approximation}
\pacs{71.10.Fd}{Lattice fermion models (Hubbard model, etc.)}

\abstract{
The time evolution of the out-of-equilibrium Mott insulator is investigated numerically
through calculations of space-time resolved density and entropy profiles resulting from the
release of a gas of ultracold fermionic atoms from an optical trap. For adiabatic,
moderate and sudden switching-off of the trapping potential, the out-of-equilibrium dynamics 
of the Mott insulator is found to differ profoundly from that of the band insulator and the metallic phase, displaying a self-induced stability that is
robust within a wide range of densities, system sizes and interaction strengths. The connection between the entanglement entropy 
and changes of phase, known for equilibrium situations, is found to extend to the out-of-equilibrium regime.  
Finally, the relation between the system's long time behavior and the thermalization limit is analyzed.}
\begin{document}
\maketitle
\newcommand{\be}{\begin{equation}}
\newcommand{\ee}{\end{equation}}
\newcommand{\bea}{\begin{eqnarray}}
\newcommand{\eea}{\end{eqnarray}}
\newcommand{\bi}{\bibitem}
\newcommand{\la}{\langle}
\newcommand{\ra}{\rangle}
\newcommand{\ua}{\uparrow}
\newcommand{\da}{\downarrow}
\renewcommand{\r}{({\bf r})}
\newcommand{\rp}{({\bf r'})}
\newcommand{\rpp}{({\bf r''})}
\section{Introduction}
The experimental realization of ultracold gases of fermionic atoms in optical
lattices is one of the major scientific breakthroughs of the past years
\cite{coldatoms1,coldatoms2,coldatoms3}. The high tunability of parameters in optical lattices permits
to study fermionic atoms with repulsive as well as attractive interactions \cite{Lahaye_reviewarticle}.
These investigations reveal a multitude of scenarios which depend on the strength and the sign
of the inter-particle interactions (for example, for the attractive case, at low temperatures, a 
complex phase diagram results, with several competing phases \cite{Koga}).

In this paper, we will consider repulsive fermions.
In addition to the Pauli exclusion
principle, the physics of such systems is governed by three distinct energy scales: the kinetic energy
of the fermions, the potential energy due to the confining trap potential, and
the fermion-fermion interaction energy.

Various numerical \cite{plateaus1,plateaus2,plateaus3,plateaus4}
and analytical \cite{plateaus5} techniques predict that this interplay gives
rise to a characteristic spatially varying density profile, displaying
coexistence of metallic, Mott-insulator and band-insulator-like regions in
different parts of the trap. Very recently, evidence for such phase-separated
density profiles in three-dimensional fermion gases has been obtained
experimentally \cite{coldatoms2,plateaus4}.

Most such investigations have been directed at {\em stationary states},
to make contact with possible ground
states of strongly-correlated,  many-electron,  condensed-matter systems. 
Trapped fermions on an optical lattice, 
however, also
allow one to study the {\em time evolution} of such systems, much more directly
and easily than in solid-state experiments, and in great detail
\cite{dyn1,dyn2,dyn3,dyn6,dyn7,dyn7b}. Very recently, for example, experiments have probed possible metastable states 
of cold atom gases, and the possibility of a dynamical tuning of the lattice and interaction parameters \cite{dyn8}. In other work \cite{turbu}, self-induced shape-stability was observed for a expanding turbulent bosonic cloud.

Motivated by such experiments, we here study numerically the time evolution of the Mott insulator, band-insulator and metallic phases after rapid, moderate and adiabatic switching-off of the trapping potential.
This allows us to address a fundamental question of many-body physics:
{\em How does the time evolution of a Mott insulator differ from that of a band insulator and of a metallic phase ?}

Before describing our methods and results, we recall that in a completely
different part of physics a similar shift from static to time-dependent (TD)
investigations is taking place: the study of entanglement in many-body systems.
Entanglement in such systems is commonly studied in connection to quantum criticality, where a deep connection between extrema of the entanglement entropy (EE) and quantum-phase transitions was found \cite{qcp1,qcp2,qcp3,qcp4}. The {\em time evolution of entanglement} has received attention \cite{tdent1,tdent2,tdent3,tdent4,tdent5,tdent6}
{\em e.g.} in the context of adiabatic quantum computation, but numerical
studies typically consider only the very particular dynamics after a quantum quench, and focus on bosons or pure spins. Very little is known about entanglement in out-of-equilibrium many-fermion states, and its possible connection to dynamic changes of phase. To shed light on these issues we here calculate the EE of the expanding cloud in parallel with its density profile.

\section{Methodology} Our Hamiltonian
$\hat{H}(\tau)= \hat{H}_0+\hat{V}(\tau)$ for the trapped fermions is 
\begin{eqnarray}
&&\hat{H}_0=-t\sum_{\langle ij\rangle,\sigma} c_{i\sigma}^\dagger c_{j\sigma}
+U\sum_i \hat{n}_{i\uparrow} \hat{n}_{i\downarrow}
+ \sum_i v_i\hat{n}_i,
\label{humo1}\\
&&\hat{V}(\tau) = - a(\tau) \sum_i v_i  \hat{n}_i .
\label{humo2}
\end{eqnarray}
In Eq. (\ref{humo1}), describing a 1D Hubbard model within
an harmonic trap, $\langle ij\rangle$ denotes nearest neighbor sites and
$\hat{n}_{i\sigma}=c_{i\sigma}^\dagger c_{i\sigma}$, with
$\sigma=\uparrow,\downarrow$, is the local density operator expressed in
terms of fermionic creation and annihilation operators. $U$ is the on-site
interaction and $t$ the inter-site hopping (below taken to be the unit of
energy). The operator $\hat{V}(\tau)$ in Eq.~(\ref{humo2}),
where  $\hat{n}_i =\sum_\sigma \hat{n}_{i\sigma}$, controls the
switching-off of the parabolic potential $v_i=k i^2/2$, via the amplitude
$a(\tau)$, with the temporal boundary conditions $a(0)=0$ for
the static trap, and $a(\tau\to\infty)=1$ for the
completely switched-off trap. Our choice for $a(\tau)$ is
\begin{eqnarray}
a(\tau) =\theta(\tau)\left[\theta(\tau_0-\tau)\sin\left(\frac{\pi}{2}\frac{\tau}{\tau_0}\right)+\theta(\tau-\tau_0)\right].
\end{eqnarray}

The rate of the switch-off of the trap is thus determined  by $\tau_0$. 
We consider three cases: $\tau_0=0^+$, $\tau_0=300$ and $\tau_0=460$, representing sudden, intermediate and adiabatic removals of the trap, respectively, in units of inverse hopping. The adiabatic case was chosen such that at any point in time the time-evolved density would be approximately equal to the ground-state density of the instantaneous potential.

To mimic the expansion of the fermion gas in absence of boundary effects, we considered a large cluster with $L=100$ sites. Furthermore, we imposed periodic boundary conditions to avoid reflections;
in this way, the ground state density of the ring without parabolic confinement is constant. We have verified, by studying larger systems, that our conclusions for the nontrivial part of the gas expansion
are not affected by finite size effects.

The ground-state density profile of the Hamiltonian $\hat{H}_0$ is obtained by solving self-consistently the single-particle Kohn-Sham (KS) equations, 
\begin{equation}
\left ( \hat{T} + \hat{v}_{KS} \right ) \varphi _i = \epsilon _i 
\varphi _i ,
\label{kohn-sham_equations}
\end{equation}
where $\hat{T}$ is the kinetic energy, $\varphi _i$ is the i-th KS orbital, $\epsilon _i$ is the i-th KS eigenvalue, and $\hat{v}_{KS} = \hat{v}_H + \hat{v}_{xc} + \hat{v}_{ext}$ is the effective single-particle potential, containing the Hartree potential $\hat{v}_H (i)= \frac{1}{2} U \hat{n}_i$, the exchange-correlation potential $\hat{v}_{xc} (i) $, and the external potential, $ \hat{v}_{ext} (i) = \sum_i v_i\hat{n}_i$ as above. The ground state density is obtained using $n_i = \sum _\kappa^{occ} | \varphi _\kappa (i) |^2$.

To obtain $v_{xc}$, we use the Bethe-Ansatz (BA) local-density approximation (LDA)\cite{lsoc}, and the ground state is obtained using lattice-density-functional theory \cite{gsn}. 
From the ground state we generate the time evolution within time-dependent density-functional theory (TDDFT)\cite{rg84}, the lattice version of which was introduced in \cite{verdozzi} for the spin-independent case and makes use of a spin-compensated, adiabatic BA-LDA. To this end, we solve the time-dependent Kohn-Sham equations, 
\begin{equation}
\left ( \hat{T} + \hat{v}_{KS} (t) \right ) \varphi _i (\tau) = i \partial _t \varphi _i (\tau)
\end{equation}
where $\hat{v}_{KS} (t) = \hat{v}_H (t) + \hat{v}_{xc} (t) + \hat{v}_{ext} (t)$
using a predictor-corrector,
split-operator algorithm, with the on-site effective potential computed in the mid-point approximation. The time-dependent density is obtained using $n_i (\tau) = \sum _\kappa ^{occ} | \varphi _\kappa (i,\tau)|^2$. Numerical convergence was checked by halving the timestep $\Delta$.

\begin{figure}
\includegraphics[width=80mm,angle=-0]{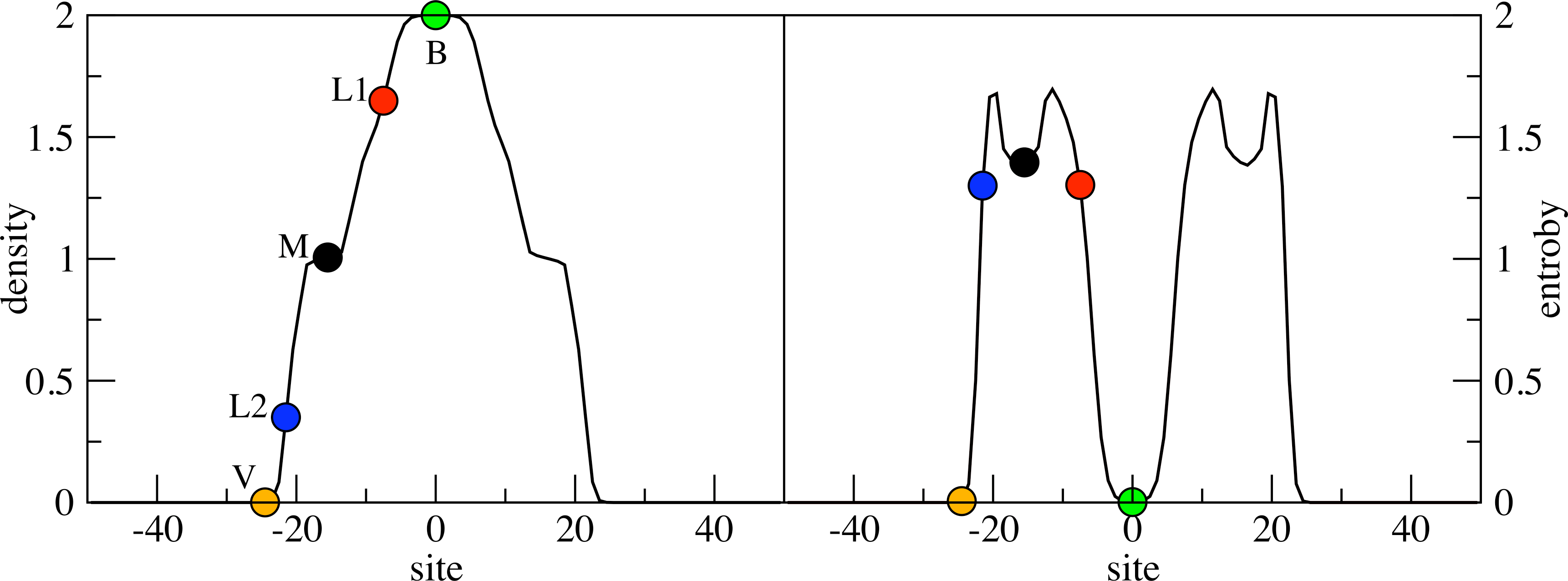}
\caption{\label{parabola} (Color online) Density (left) and entanglement  entropy
(right) profiles for a chain with $L=100$ sites and periodic boundary
conditions, $N=60$ spin compensated fermions with on-site interaction $U/t=8$, trapped by a static parabolic potential of
curvature $k/t=0.05$. The colored symbols represent
sites in the band insulating (B), Luttinger liquid (L1, L2), Mott
insulating (M) and near-vacuum (V) regions. In panels b) and c) of Figs. \ref{adia},\ref{interm},\ref{step} below,
the same colors refer to the same sites.}
\end{figure}

The $xc$ potential obtained in \cite{lsoc} is a discontinuous function of the density at half filling \cite{mottepl}.
During the ALDA dynamics, this discontinuity in $v_{xc}$, depending itself on the TD density, 
makes the time evolution numerically
challenging \cite{verdozzi, vcu}. To make
the problem tractable, we slightly smoothened the
discontinuity and used a recursive time step in the time propagation,
to ensure that between $\tau$ and $\tau+\Delta$ no jump in
$v_{xc}$ was missed. For a smoothed $v_{xc}$, the original discontinuity
broadens over a range $\delta n \approx 0.095$, and its value is reduced by $\approx 10\%$. Due to this, the shape of the Mott plateaus gets slightly rounded (see Fig.~\ref{parabola}). 
This does not affect the essence of our findings:  we have verified that, on reducing the smoothing, the Mott physics we
address below becomes in fact  more pronounced.
\section{Entanglement and TDDFT}
The EE of the homogeneous one-dimensional Hubbard model is given,
as a function of filling $n$ and interaction $U$, by the expression
\cite{qcp3,francacapelle,gu_entanglement} 
\begin{eqnarray}
\label{entafor}
\mathcal{E}(n, U)= &-&2\left(\frac{n}{2}-\frac {\partial e(n,U)}{\partial U} \right) \log_2 \left [\frac{n}{2}-\frac {\partial e(n,U)}{\partial U}\right]\nonumber\\
&-&\left(1-n+\frac {\partial e(n,U)}{\partial U} \right) \log_2 \left [1-n+\frac {\partial e(n,U)}{\partial U}\right]\nonumber\\
&-&\frac {\partial e(n,U)}{\partial U}  \log_2 \left [\frac {\partial e(n,U)}{\partial U}\right] , 
\label{eq_entanglement}
\end{eqnarray}
where $e(n,U)$ is the per-site ground-state energy. In order to evaluate this expression
in the present case we use a parameterization of $e(n,U)$ introduced in \cite{lsoc}, according to which
\begin{eqnarray}
\label{ederiv1}
\frac{\partial e(n,U)}{\partial U}=\frac{2}{\pi} \frac{\partial\beta}{\partial U}\left[ \frac{\pi n}{\beta} \cos \frac{\pi n}{\beta}- \sin \frac{\pi n}{\beta} \right],
\end{eqnarray}
\begin{eqnarray}
\frac{\partial\beta}{\partial U} = \frac{ \frac{\pi}{4}  \int_{0}^{\infty} dx\frac{J_0(x)J_1(x)}{\cosh^2(Ux/4)}    }{\left[ \frac{\pi}{\beta} \cos \frac{\pi}{\beta}- \sin \frac{\pi}{\beta} \right]},
\label{ederiv2}
\end{eqnarray}
and where $\beta$ is determined  from the exact energy density of the 1D homogeneous Hubbard model at half-filling, 
i.e. $-\frac{2\beta}{\pi}\sin(\frac{\pi}{\beta})=-4\int_0^\infty dx \frac{J_0(x)J_1(x)}{[x(1+ exp(Ux/2)]}$.
\noindent In the present spatially inhomogeneous case, we evaluate the per-site 
entropy in terms of the spatially varying per-site density, $n_i$, which amounts
to making a local-density approximation to the entanglement \cite{francacapelle}. 
The inhomogeneous
density profile itself is obtained from the adiabatic local-density approximation (ALDA) to 
the time-dependent lattice-density-functional theory \cite{verdozzi}, using the same parameterization of the 
Bethe-Ansatz solution \cite{lsoc}, but solving the time-dependent (TD) Kohn-Sham equations \cite{rg84}
on the lattice \cite{verdozzi} instead of the stationary ones \cite{gsn}.

These static and TD LDA-like approaches for the Hubbard model are described in more detail in \cite{lsoc,verdozzi,mottepl,francacapelle}, where they have been tested and
benchmarked against exact diagonalization, density-matrix renormalization
and quantum Monte Carlo calculations and shown to attain an accuracy of the order of a few percent for energies,
particle densities and entropies. Their favorable computational cost permits time-resolved studies of systems of hundreds of sites for
any boundary condition, even in the absence of simplifying symmetries.

\begin{figure*}[tbh]
\includegraphics[width=160mm,angle=-0]{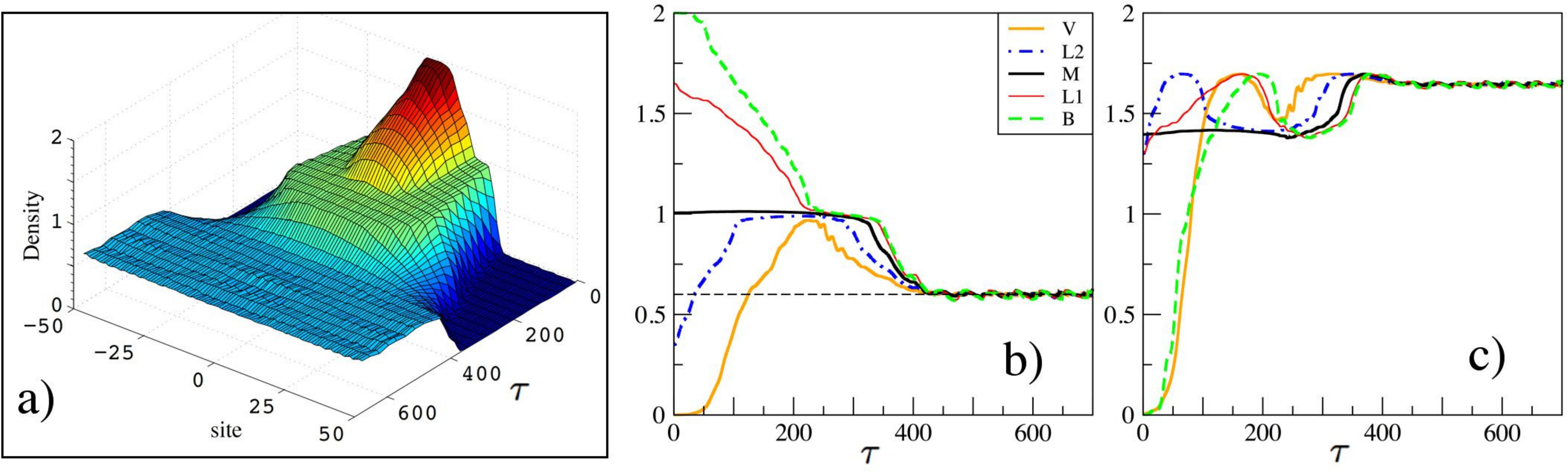}
\caption{\label{adia} (Color online) Panel a): Time and space resolved density profile for adiabatic switching-off of the trapping potential. At time $\tau=0$ the
density profile is that of Fig.~\ref{parabola}, while for later times
the curvature of the trap is slowly reduced, allowing the density profile
to expand. Panel b): Cross sections of panel a), showing the time evolution
of the density at the representative sites indicated in fig.~\ref{parabola}. The thin horizontal line indicates the uniform density distribution $n_0$ ($=0.6$) for the untrapped system.
Panel c): Time evolution of the entanglement  entropy at the same sites. For the EE, the band insulator phase (green curve) and the vacuum one (orange curve) closely resemble each other.  This is a reflection of the particle-hole symmetry, which is contained in the definition for the entropy, Eq. (\ref{eq_entanglement}). The agreement is not perfect however, since the original ground state fulfills this symmetry only approximately.}
\end{figure*}

\section{Results and Discussion}
We start this section  with a brief analysis of the ground-state properties of our system(s).
Figure~\ref{parabola} shows a representative ground-state density profile
and entanglement-entropy profile. Mott insulating
(M), band insulating (B) and vacuum (V) regions correspond to flat regions
in the density profile and local minima in the entanglement  entropy profile. These minima
are separated by metallic regions (L1, L2), which in one dimension display Luttinger
liquid phenomenology. We now adopt this representative density
profile as the initial state for the subsequent time evolution with
the full Hamiltonian $\hat{H}(\tau)= \hat{H}_0+\hat{V}(\tau)$.
%
%

Panel a) of Fig.~\ref{adia} illustrates the time-and-space resolved
density profile resulting from a very slow ( i.e., a "numerically adiabatic") switching-off of the trap. 
For such a slow perturbation, and before the particles reach the boundaries,
the results can be
interpreted in terms of  trapped-equilibrium-systems considerations \cite{plateaus1,plateaus5}
(see below). However, in the following we find useful to adopt
a time-dependent perspective, which remains appropriate also for faster switching-off of the trap.

At time $\tau=0$ the initial density profile is that of
Fig.~\ref{parabola}. As expected for adiabatic switching, for very long
times the density evolves towards the ground state of the unconfined system,
which in our case corresponds to a uniform distribution of $60$ fermions
over $100$ sites on a ring.

Panel b) shows explicitly the time evolution of the five
representative sites.
All sites ultimately attain this density, but in very different ways.
The metallic regions L1 and L2 
start evolving towards $n_0$ as soon as the trap is
reduced. Similarly, the vacuum region (V) gets filled up almost
immediately. However, at intermediate times, roughly between $\tau=100$ and $\tau=400$, the vacuum receives
more fermions than would correspond to the uniform final state.

The densities in the Mott regions (M), on the other hand, maintain their  $\tau=0$ value until $\tau \approx 350$, i.e.
a persistency of the Mott phase is observed:  This is consistent with previous work \cite{dyn6,dyn7,tdent6}.
Furthermore, when the density at the originally metallic sites reaches 1 from above (L1) or below (L2), it, too, develops the
characteristic Mott behavior and persists for an extended period
of time. The corresponding transient flat regions are clearly visible in the curves labelled L1 and L2 in panel b). 
Differently from the Mott phase, the band insulator
starts evolving towards the uniform state as soon as the trap begins to be reduced. This points at a basic difference between band and Mott insulators: 
The Mott insulator self-stabilizes due to particle-particle interactions, while the band insulator
requires an external potential to be stable. Panel c) shows the time evolution of the
EE. In contrast with the density results, the entanglement curves for vacuum and
band insulating regions display very similar behavior.
From an equilibrium-regime perspective, the above results  can be rationalized
in terms of energetics arguments \cite{plateaus1,plateaus5}: in strong traps 
band and Mott insulators coexist with compressible domains. As the trap 
curvature is reduced, the band insulator becomes energetically unfavorable, 
while the Mott insulator is sustained due to the rigidity arising from the discontinuity in $v_{xc}$. Also, the vacuum and band insulating regions, which have very different density profiles, display very similar EE behaviour;
this is because the  EE
is related to the degrees of freedom that are available for storing or 
recovering information, and this number is zero if a site is completely filled
or empty.

\begin{figure*}
\includegraphics[width=160mm,angle=-0]{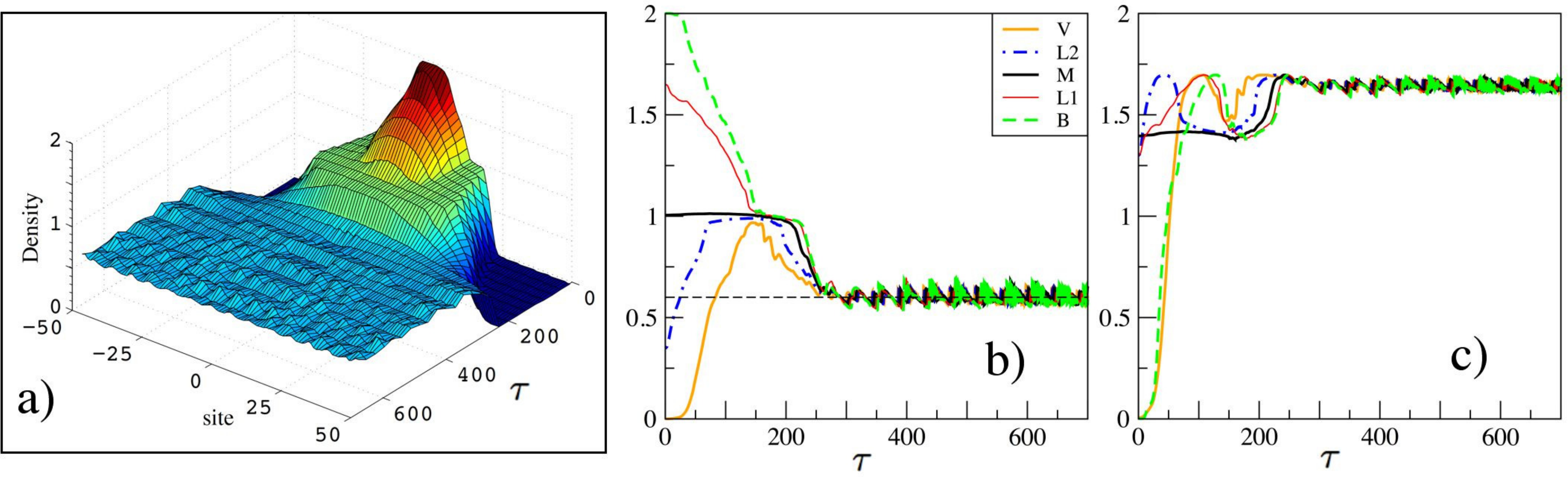}
\caption{\label{interm} (Color online) As Fig.~\ref{adia}, but for moderate switching-off of the trap.}
\end{figure*}
%
%
%
Results for moderate switching-off
of the trap are in Figure~\ref{interm}. 
The overall features of the adiabatic curves are preserved, but compressed in a shorter timescale.
For example, for long times the entanglement  entropy evolves towards that of the
uniform system, but again with significant delay for the Mott insulator, which exhibits resistance against melting until about $\tau=250$. 
We note, however, also some differences from Fig. \ref{adia}, e.g. for long times the density (and the EE)
oscillates around the uniform state. 
Unlike the total particle number, the total EE is not conserved, and reaches, not necessarily in a monotonic way, 
a maximum at long times. In equilibrium, extrema of the EE are known \cite{qcp1,qcp2,qcp3,qcp4} 
to be markers of quantum phase transitions.
In Figs. \ref{adia},\ref{interm}, the transitions from a Luttinger liquid to a Mott insulator and from Mott to Luttinger correspond to extrema in the EE. 
In principle, such correspondence
might be spoiled away from equilibrium. Our simulations show, however, that this is not necessarily the case.

\begin{figure*}
\includegraphics[width=160mm,angle=-0]{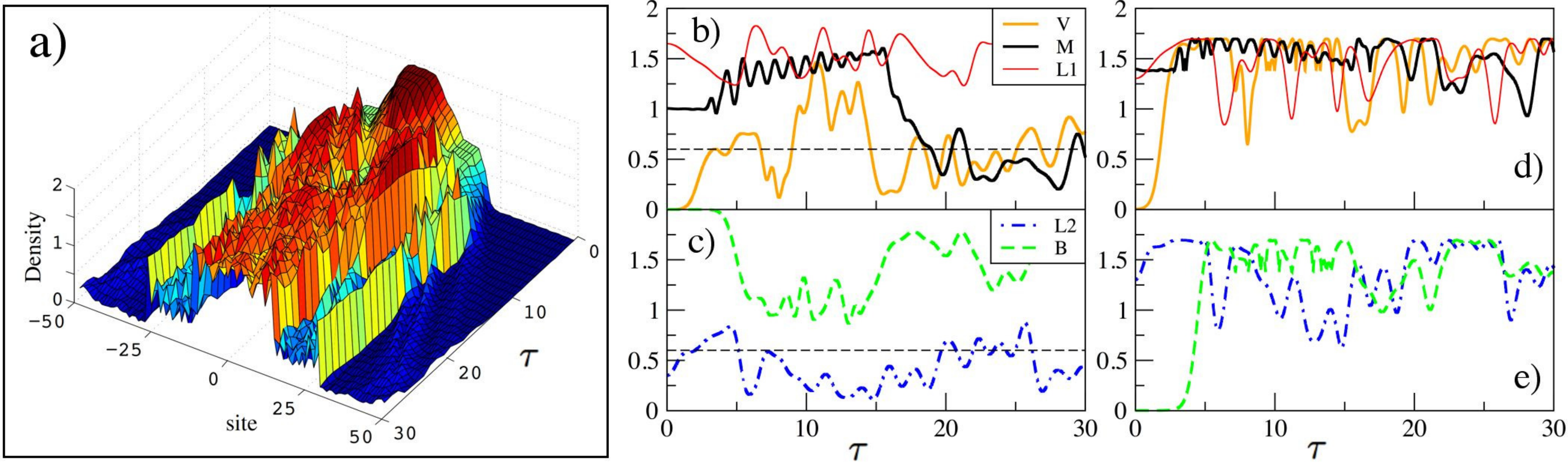}
\caption{\label{step} (Color online) As Fig.~\ref{adia}, but for instantaneous switching-off
of the trap.}
\end{figure*}

%
%
%
Finally, Fig.~\ref{step} refers to instantaneous switching.
Both density and entanglement  entropy show strong oscillations on a short time scale.
Most likely, such oscillations are an artifact of our ALDA ; indirect evidence for this also comes from tDMRG studies of
sudden quenches of the confining potential, which show a smooth expansion of the density profiles
\cite{dyn7, tdent6, NJP}.
The main aim of Fig.~\ref{step} is to show an instance
where the ALDA (but not lattice TDDFT \cite{verdozzi}) scheme becomes inadequate. By contrast, 
the lattice TDDFT-ALDA \cite{verdozzi} should be
useful to follow the long-time evolution for slow and moderately fast switching off, i.e. situations
which currently are not easily accessible within tDMRG calculations.

\begin{figure}[h]
\hspace{+0.0cm}
\includegraphics[width=80mm,angle=-0]{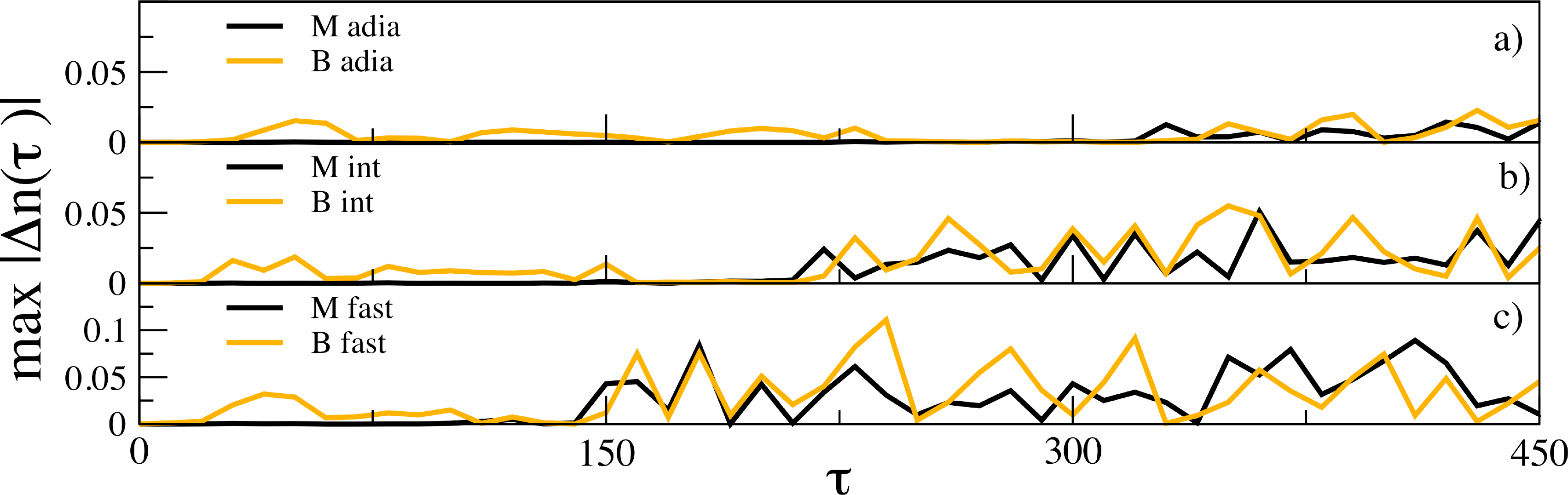}
\caption{\label{selfcon} Maximum absolute difference $\Max_{1 \le i\le L} | n_i (\tau)-n^{gs}_i(\tau)|$ between 
time evolved densities and those from the instantaneous ground state 
of $\hat{H}(\tau)$, at the band insulator (B) and Mott plateaus (M) points (see Fig. \ref{parabola}). 
Top to bottom panels: adiabatic ($\tau _0 = 460$, as in Fig.\ref{adia}), 
intermediate ($\tau _0 = 300$, as in Fig.\ref{interm})  and faster removal of the trap ($\tau _0 = 200$).}
\end{figure}

It is useful, at this stage, to quantify 
of the degree of adiabaticity in our results. To this end, we show in Fig. \ref{selfcon}
three switching-off speeds. At each time $\tau$, we considered 
the maximum difference (among all sites) between the time evolved 
densities and those obtained by the instantaneous ground state of the Hamiltonian. 
One sees that the TD results of  Fig. \ref{adia}  are  quite close to the instantaneous 
ground states ones. On the other hand, for faster perturbations, 
the differences are more noticeable, a sign of significant 
departure from adiabaticity. 

We stress that these findings are not related to a particular
choice of the system parameters. We observed (not shown) the same behaviour  in simulations with
different amplitude modulations, numbers of fermions and lattice sites,
different values for the on-site interaction, etc. \cite{2Dexpand}.
\section{Thermalization, ground state, and TDDFT} \label{HEAL}
Our time-evolution results permit to address two other interesting issues within TDDFT, namely the 
system's thermalization (i.e. the achievement of local equilibrium through interactions among the particles)
and if it possible or not for the system to reach the ground state, once the trap is removed.

Generally,  achieving thermalization or 
attaining the ground state are distinct processes:
In general, after a parameter quench,  in a finite system a
ground state is not reached without exchange of energy; by contrast, thermalization  
is possible also when the system remains isolated after the quench. Thus, even for our zero-temperature calculations, 
thermalization remains a meaningful concept. As an indicator of local equilibrium in the long-time limit we can use 
the average value of a suitable one-body operator. In the
case of TDDFT, the one-particle density $n_i(\tau)$ is a natural
choice.

For finite isolated systems, exact diagonalization studies have shown that,
under quite general conditions, thermalization occurs \cite{rigol_nature}. 
On the other hand, experimental results for 1D interacting bosons \cite{Kinoshita} and
several theoretical studies  \cite{Cazalilla, Manmana, Biroli, Kronenwett}
indicate that, in some cases, the quasi-stationary states after an interaction quench can be non-thermal.
Overall, it is fair to say that, at present, the issue of thermalization in the presence of a global quench is not completely settled yet. 

In our present context, these generic remarks suggest the following specific questions from a TDDFT perspective:
i)  How does thermalization occur in our system, when the confining potential is removed?
ii) What is the relation between the state reached by our system in the long time regime and the ground state of the final Hamiltonian? 
To briefly  address these points,  let us first consider for definiteness the exact many-body dynamics of our system, Eq.(\ref{humo2}), for two kinds of perturbations: i) sudden and ii) adiabatic. 
In the initial ground state $|g\rangle$ with energy $E_g$, the parabolic potential $\hat{V}_P=\sum_i v_i \hat{n}_i$ contributes a positive energy $\langle g| \hat{V}_P |g\rangle$. After a sudden removal of the parabolic trap (as in Fig.\ref{step}), the system has a new, time independent Hamiltonian $\hat{H}'$, but the same initial state $|g\rangle$ and the (initial) average energy $E'= \langle g|\hat{H}'|g \rangle = E_g- \langle 
g|\hat{V}_P| g\rangle$  is conserved at all times. Since, in general, $E' \neq E'_{g'}=\langle g'|\hat{H}'|g'\rangle$, 
our system may thermalize, but cannot reach the ground state.
On the other hand, if no symmetry restriction apply, when energy is removed from the system continuously and infinitely slowly (adiabatic switch-off of $\hat{V}_P$, as in Fig. \ref{adia}), the ground state can be reached.
For "intermediate-speed" perturbations, as in Fig.\ref{interm}, in general thermalization may occur,
even if the system does not reach the ground state.

In an exact TDDFT description, the exact TD density is accessed. Hence, the considerations above about thermalization and/or reaching the ground state still hold.  However, the adiabatic BA-LDA used here is a local approximation in space and time, and thus dissipative effects are neglected. Consequently, the possibility that the system's thermalization is described incorrectly cannot be ruled out. For
example, within the simulation interval considered, the results of Fig. \ref{step} show no indication that a uniform density (indicative of thermalization within TDDFT) 
is going to be established. On the other hand, for a very slow (adiabatic in a numerical sense) removal of the trap, such as in Fig.\ref{adia}, our treatment is expected to be quite accurate in describing the way the system approaches the ground state.
\section{Conclusions}
In summary, our simulations show that
(i) the time evolution of the expanding cloud  displays a wide variety of non-equilibrium phenomena (overshooting, transients, self-induced stability, etc), all of which should be experimentally accessible with todays technology. In particular, optical experiments can be used to investigate the time evolution of the Mott insulator  -- a state of matter that in ordinary condensed-matter situations can only be studied in static situations;
(ii) a connection between the entanglement entropy and phase changes is observed also in non-equilibrium situations, thus suggesting the possibility of investigations and applications of quantum information concepts in dynamical settings;  
and (iii) if accurate enough potentials are used, TDDFT is a useful tool for characterizing and analyzing the long-time behavior of the expanding cloud, and to describe phenomena such as the approach to the ground state or the  thermalization of initial states that are far from equilibrium.
\acknowledgments
CV is supported by ETSF (INFRA-2007-211956).
KC is supported by FAPESP and CNPq.

\end{document}